\documentclass[twocolumn,prl]{revtex4}
\usepackage{epsfig,amssymb,amsmath}

\begin{document}
\title{Breathing Charge Density Waves in Intrinsic Josephson Junctions}

\author{Yu. M. Shukrinov~$^{1}$}
\author{H. Abdelhafiz~$^{1,2}$}

\address{$^{1}$ BLTP, JINR, Dubna, Moscow Region, 141980, Russia}
\address{$^{2}$Nile University, Smart Village, Cairo, Egypt}

\date{\today}

\begin{abstract}
We demonstrate the creation of a charge density wave (CDW) along a stack of coupled Josephson junctions in layered superconductors. Electric charge in each superconducting layer oscillates around some average value, forming a breathing CDW.  We show the transformation of a longitudinal plasma wave to  CDW in the state corresponding to the outermost branch. Transitions between different types of CDW's related to the inner branches of current voltage characteristics are demonstrated. The effect of the external electromagnetic irradiation on the states corresponding to the inner branches differs crucially  from the case of the single Josephson junction. The Shapiro steps in the IV-characteristics of the junctions in the stack do not correspond directly to the frequency of radiation $\omega$. The system of Josephson junctions behaves like a single whole system: the Shapiro steps or their harmonics in the total IV-characteristics appear at voltage $\sum V_l=N_R\frac{m}{n}\omega$, where $V_l$ is the voltage in $l$-th junction, $N_R$ is the number of JJ in the rotating state, and $m$ and $n$ are integer numbers.
\end{abstract}
\maketitle

Charge density waves in solids are static (unchanging in time) periodic modulations of the conduction electron density. Such charge modulations are well known in different low dimensional systems, as well as cuprate superconductors \cite{gruner94,moncton77,fujita04,kivelson03,vojta09}. Recently,  the CDWs were discovered in $YBa_2Cu_3O_{6+x}$  \cite{chang12,ghiringhelli12}. It was mentioned in Ref.\cite{fradkin12} that ``this discovery places charge orders center stage with superconductivity, suggesting that they are intertwined rather than competing''.

In contrast to these classical CDWs, where the origin of CDW is related to the peculiarities of the Fermi surface \cite{gruner94}, another type of CDW  related to the non-equilibrium nature of the AC Josephson effect in HTSC is possible. Machida et al. \cite{machida99a}  demonstrated the instability of longitudinal plasma waves in the states corresponding to the inner branches and showed that there is a static CDW along the c-axis.  They suggested  that the branching of current voltage characteristics (IV-characteristics) is a result of transitions between different spatial  CDW modes.

In this Letter, we discuss the nature of different CDW spatial modes realized due to the coupling between intrinsic Josephson junctions  in HTSC. We suggest a mechanism of formation of CDWs as  states with different numbers of Josephson junctions in the rotating (R) state, and oscillating (O)   state. Actually,  CDW is not absolutely static and is, therefore, referred to here as a breathing CDW. Two different types of CDW related to the outermost branch and inner branches of the IV-characteristics are demonstrated. We stress a different origin for CDWs realized in the states corresponding to the inner branches of the IV-characteristics.

The influence of the external electromagnetic radiation on the states corresponding to the inner branches cardinally differs from that for the  outermost branch or for single JJ. We demonstrate that the Shapiro step in the IV-characteristics of each junction of the stack does not correspond to the radiation frequency directly. System of coupled JJ reacts as a single one.

To investigate the CDW in HTSC we use the one-dimensional CCJJ+DC model with the gauge-invariant phase differences $\varphi_l(t)$  between S-layers $l$ and $l+1$  described by the system of equations:
\begin{equation}
\label{syseq} \left\{\begin{array}{ll} \displaystyle\frac{\partial \varphi_{l}}{\partial
t}=V_{l}-\alpha(V_{l+1}+V_{l-1}-2V_{l})
\vspace{0.2 cm}\\
\displaystyle \frac{\partial V_{l}}{\partial t}=I-\sin \varphi_{l}-\beta\frac{\partial \varphi_{l}}{\partial t} + A\sin \omega t + I_l^{noise}
\end{array}\right.
\end{equation}
where $t$ is the dimensionless time normalized to the inverse plasma frequency $\omega^{-1}_p$ ( $\omega_{p}=\sqrt{2eI_c/\hbar C}$, $C$ is the capacitance of the junctions,  $\beta=1/\sqrt{\beta_{c}}$, $\beta_{c}$ is the McCumber parameter), $\alpha$  gives the coupling between junctions\cite{koyama96}, $\omega$ and $A$ are the frequency and amplitude of the external electromagnetic radiation, respectively.  To find the IV-characteristics of the stack of the intrinsic JJ, we solve this system of nonlinear differential equations (1) using the fourth order Runge-Kutta method. In our simulations we measure the voltage in units of $V_0=\hbar\omega_p/(2e)$, the frequency in units of $\omega_{p}$, and the bias current $I$ in units of $I_c$. Time dependence of the electric charge in the superconducting layers is investigated by  Maxwell equation $\emph{div} (\varepsilon\varepsilon_0 \vec{E}) = Q$.
The charge density $Q_l$ (in the following text referred to as charge) in the S-layer $l$ is proportional to the difference between the voltages $V_{l}$ and $V_{l+1}$ in the neighbor insulating layers $Q_l=Q_0 \alpha (V_{l+1}-V_{l})$,
where $Q_0 = \varepsilon \varepsilon _0 V_0/r_D^2$.
Numerical calculations have been done for a stack with the coupling parameter $\alpha=1$, dissipation parameter $\beta= 0.2$ and periodic boundary conditions.  Details of the model and simulation procedure can be found in Ref.\cite{shukrinov-prb07}.

\begin{figure}
 \centering
\includegraphics[width=70mm]{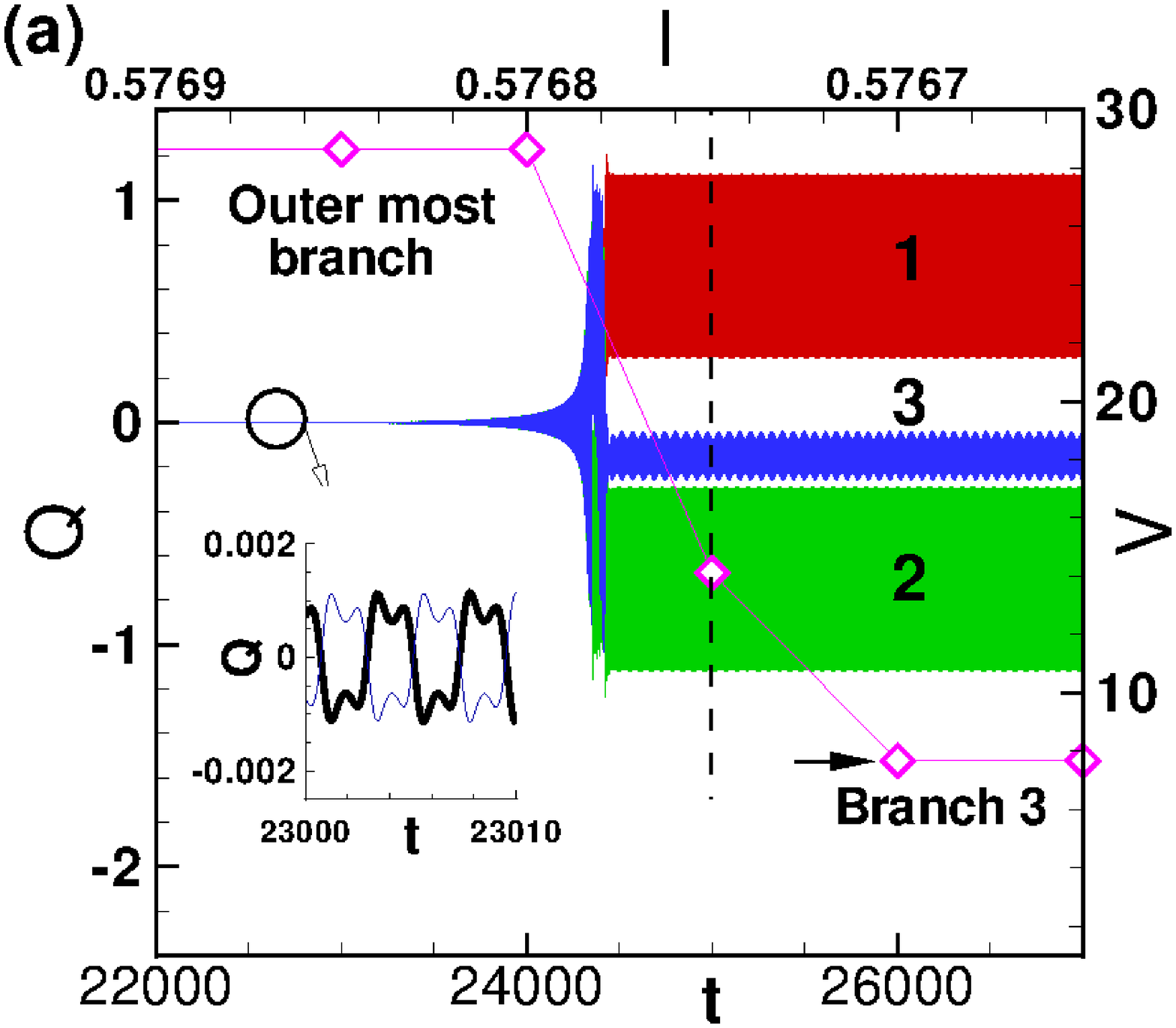}
\includegraphics[width=70mm]{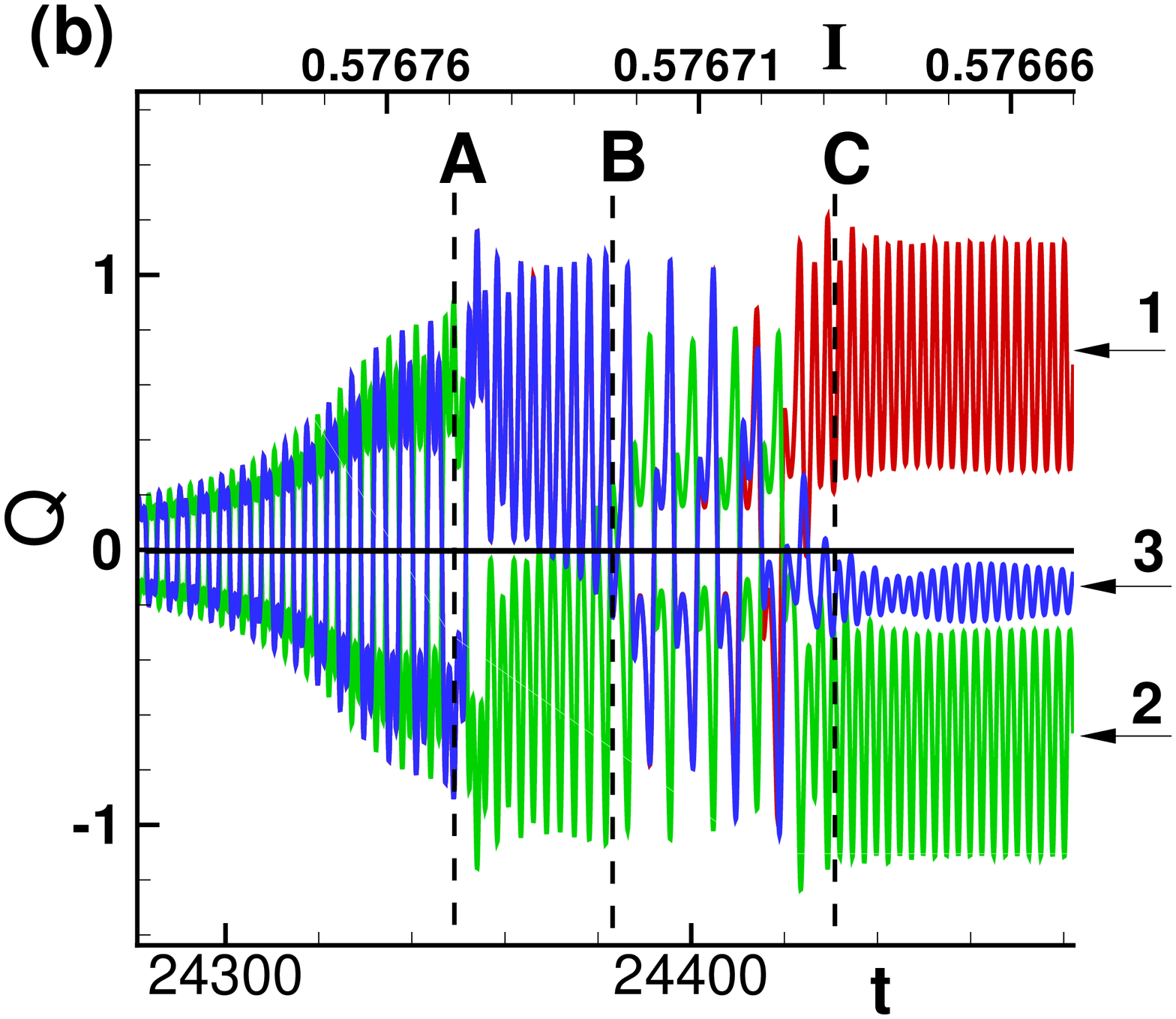}
\includegraphics[width=70mm]{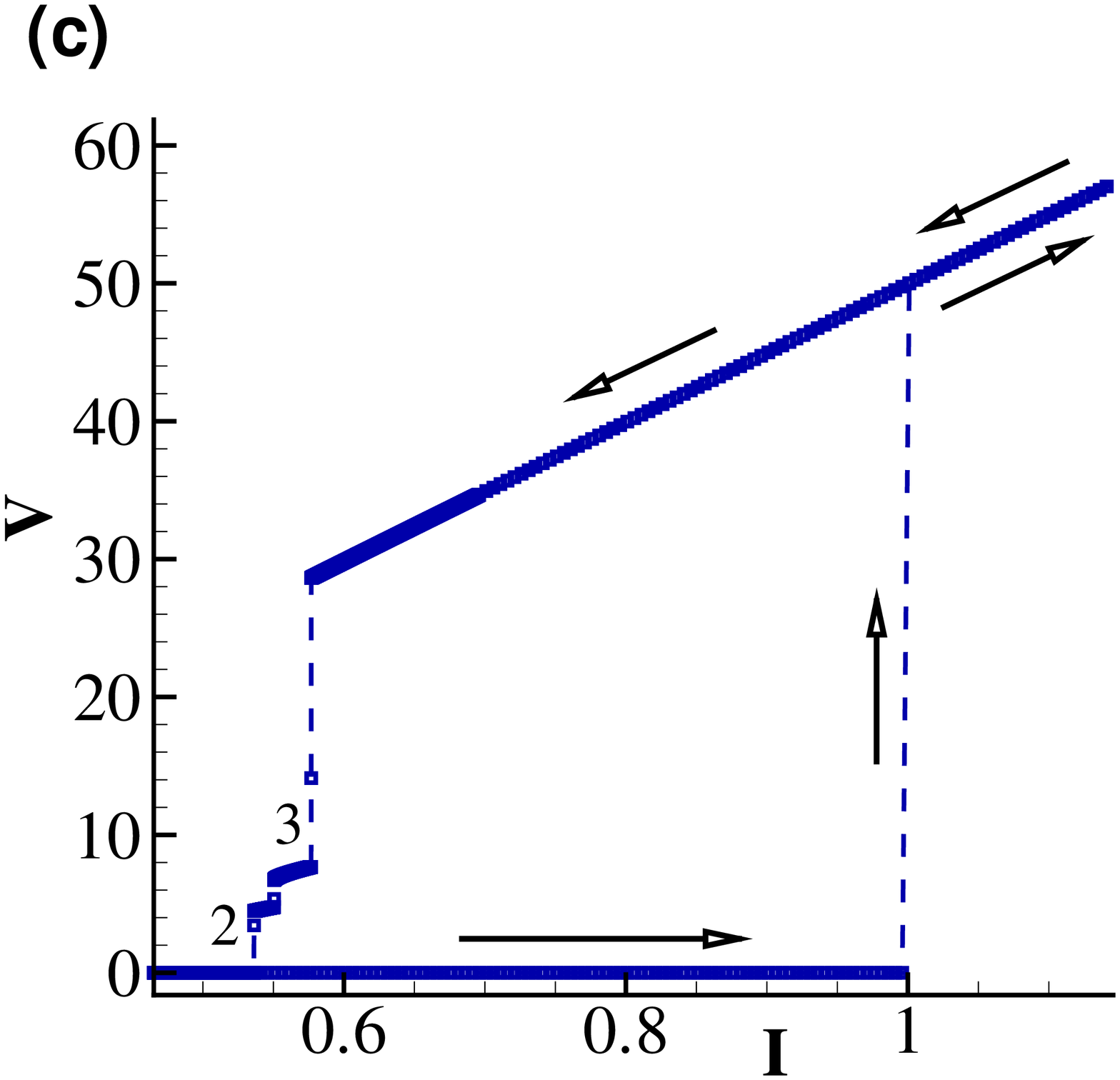}
\caption{(Color online) (a) $Q(t)$ dependence together with the IV characteristics (curve with the symbols related to the right and upper axes) near transition from the outermost branch to branch number 3 (filled arrow indicates the branch number 3). The dashed line passing through the current symbol stress that the system are in the CDW-state.  The inset shows the LPW oscillations ($\pi$-mode) in the growing region of the parametric resonance. (b) The enlarged part of the  $Q(t)$ dependence with a transition region from the LPW to the CDW. (c) One-loop IV-characteristics for the same stack. The arrows show the directions of bias current changes throughout the simulation. Branches 2 and 3 are realized in this case.}
 \label{1}
\end{figure}

Let us first consider results related to the transition from the outermost branch (all junctions in R-state) to inner branch (some junctions in O-state) \cite{matsumoto99}. Such a transition happens in the resonance region, where a  longitudinal plasma wave (LPW) with a definite wave number is created. In the growing region of the parametric resonance we observe  the charge oscillations corresponding to the $\pi$-mode of the LPW (see inset in Fig.~\ref{1}(a)).

In Fig.~\ref{1}(a) we show  the $Q(t)$ dependence together with the IV characteristics (thin curve with diamond symbols) near transition from the outermost branch  to an inner branch with three JJs in the R-state (branch number 3 in Fig.\ref{1}(c)). To simplify the picture, we show the charge oscillations in three layers only. The $Q(t)$ dependence shown,  is characterized by exponential increase of the charge in the S-layers in the parametric resonance region, some transition region, and by oscillations of different amplitude around different average values in each superconducting layer of the stack.

As we see,  branch number $3$ corresponds to the state with a CDW: the electric charge in each superconducting layer oscillates around some average value forming a breathing CDW along the stack. To stress this fact  we plot in Fig.~\ref{1}(a) the dashed line passing through the current symbol corresponding to the current value $ I=0.57675$. The transition from one branch to the other is a result of the phase dynamics of the system, and it is  not related directly to the change of the bias current value. As we see in Fig.~\ref{1}(a), the dynamical transition from the outermost branch to branch 3 happens in one time domain, i.e. at some chosen value of bias current. Intermediate values of voltage between two branches correspond practically to the formed CDW, and their values depend on the time interval which the system spent in the formed state.

Figure ~\ref{1}(b) illustrates the enlarged part of the  $Q(t)$ dependence around the transition from the LPW to the CDW state corresponding to the outermost branch. As we mentioned above, the CDW forms in the current interval with initial current value related to the outermost branch. We can distinguish four different regions here. Before  point $A$ the system is in the state with LPW. Then, in region $A-B$ we can see  the breathing CDW, when the charge in odd layers breaths around positive average value, while the charge in the even layers breaths around negative average value. The time interval the system spend in this state is very short.  In the region $B-C$ the system is in CDW state also, but the averaged values of charge in odd and even layers are decreased  compared with $A-B$ region. After point $C$ the system goes to the state with CDW related to the inner branch. Analysis shows that it is a branch  number {\it 3}  with 7 JJ in the oscillating state.

\begin{figure}
 \centering
\includegraphics[height=40mm]{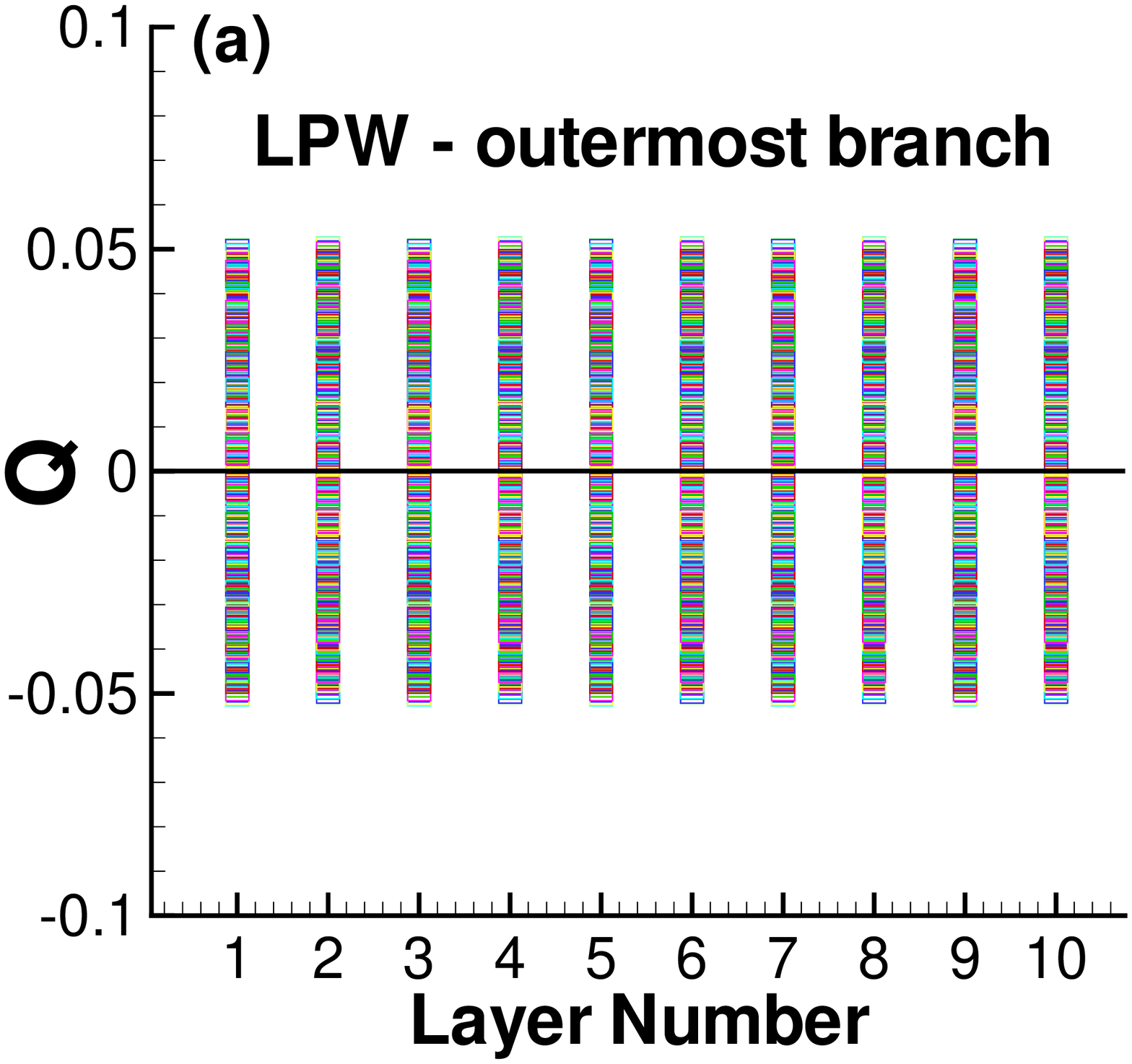}\includegraphics[height=40mm]{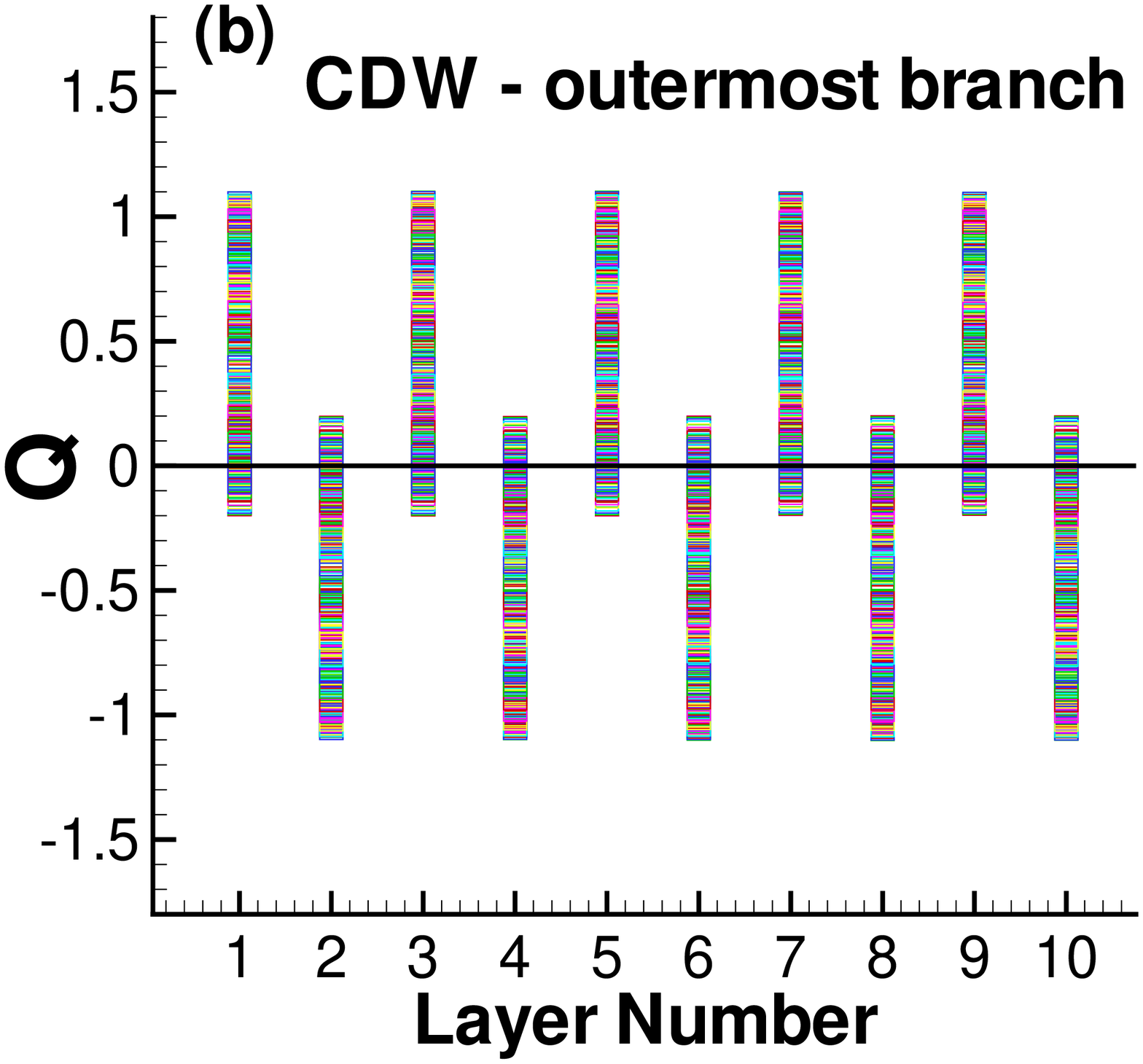}
\includegraphics[height=40mm]{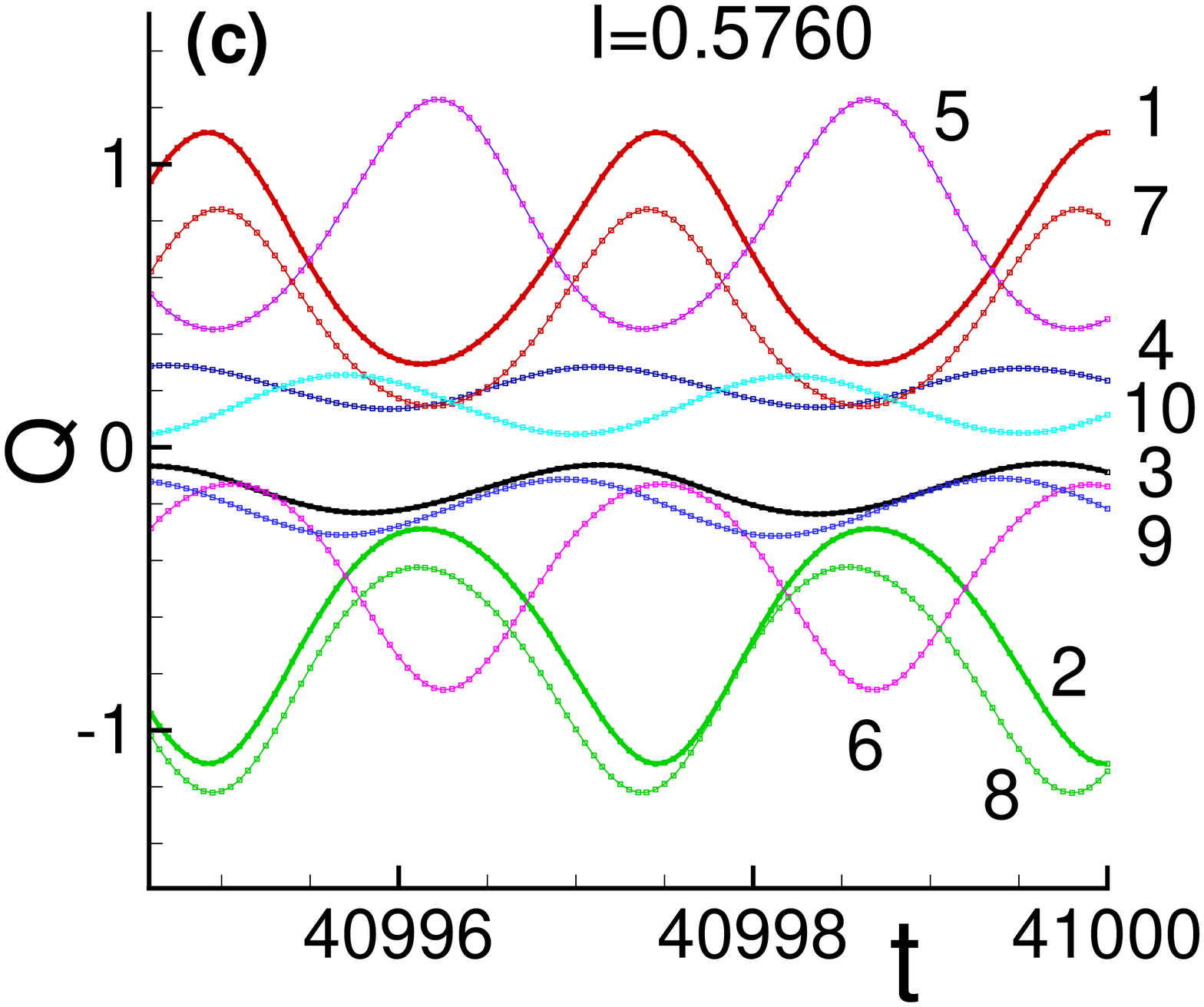}\includegraphics[height=40mm]{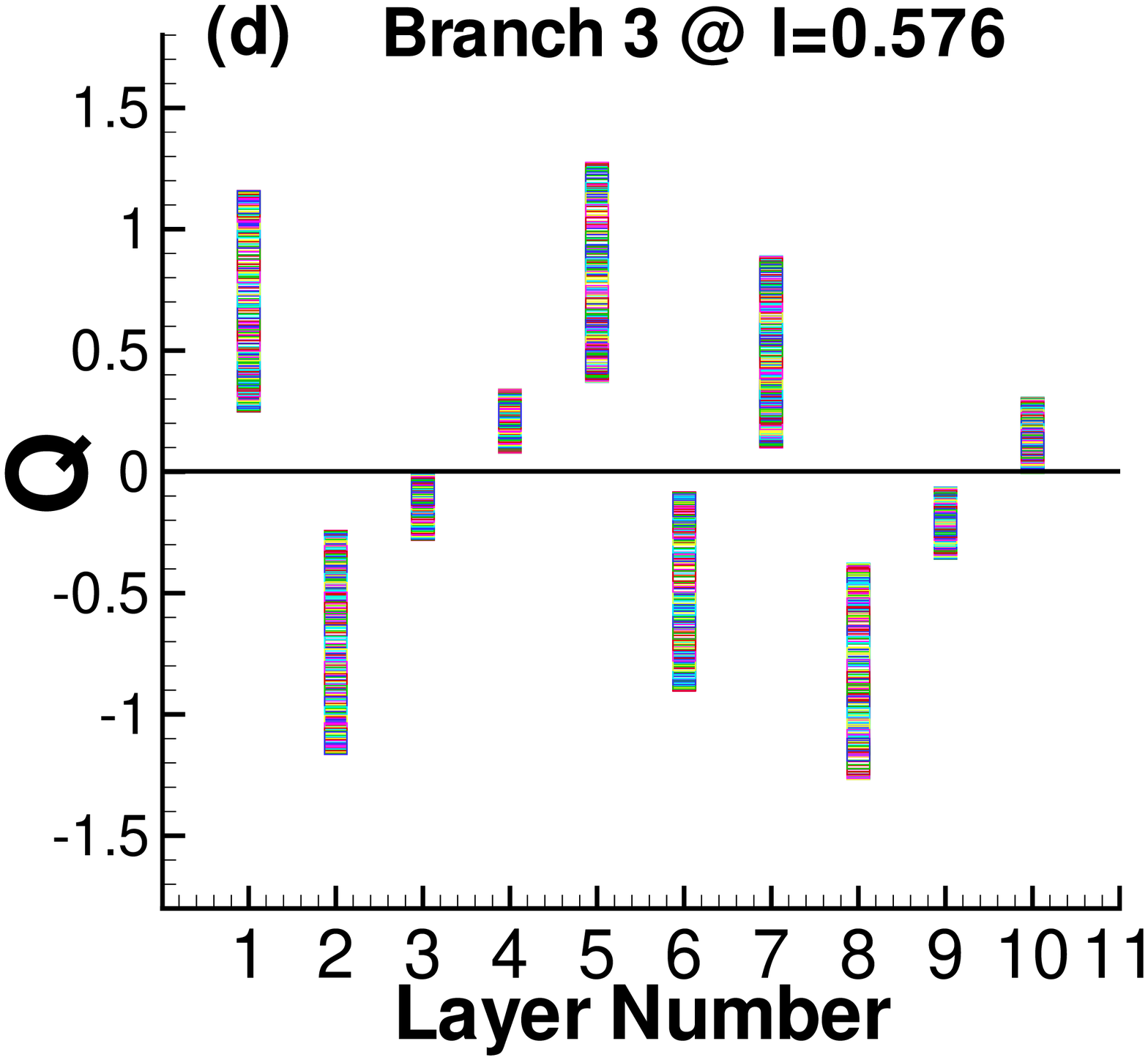}
\caption{(Color online) (a)  Charge variation  along the stack during time domain before point $A$ in Fig.~\ref{1}(b); (b) The same in the $A-B$ region; (c) Charge oscillations in different layers, corresponding to the inner branch 3; (d) The same as in (a) after point $C$.}
 \label{2}
\end{figure}

Let us now consider the charge oscillations in S-layers before the point $A$ again.  If we plot  the values of charge in each layer in time, we  get the distribution of charge along the stack presented in  Fig.~\ref{2}(a). The charge distribution in the $A-B$ region is shown in Fig.~\ref{2}(b). We see clearly the LPW in the first case and the breathing CDW in the second one.

The charge oscillations in different layers in the CDW state, corresponding to the inner branch 3 are shown in Fig.~\ref{2}(c). We see that in the four S-layers (3,4,9,10) the value of charge is much smaller compared with other layers. If we take into account all charge values in time at $I=0.576$, we obtain the picture presented in Fig.~\ref{2}(d).

\begin{figure}
 \centering
\includegraphics[height=40mm]{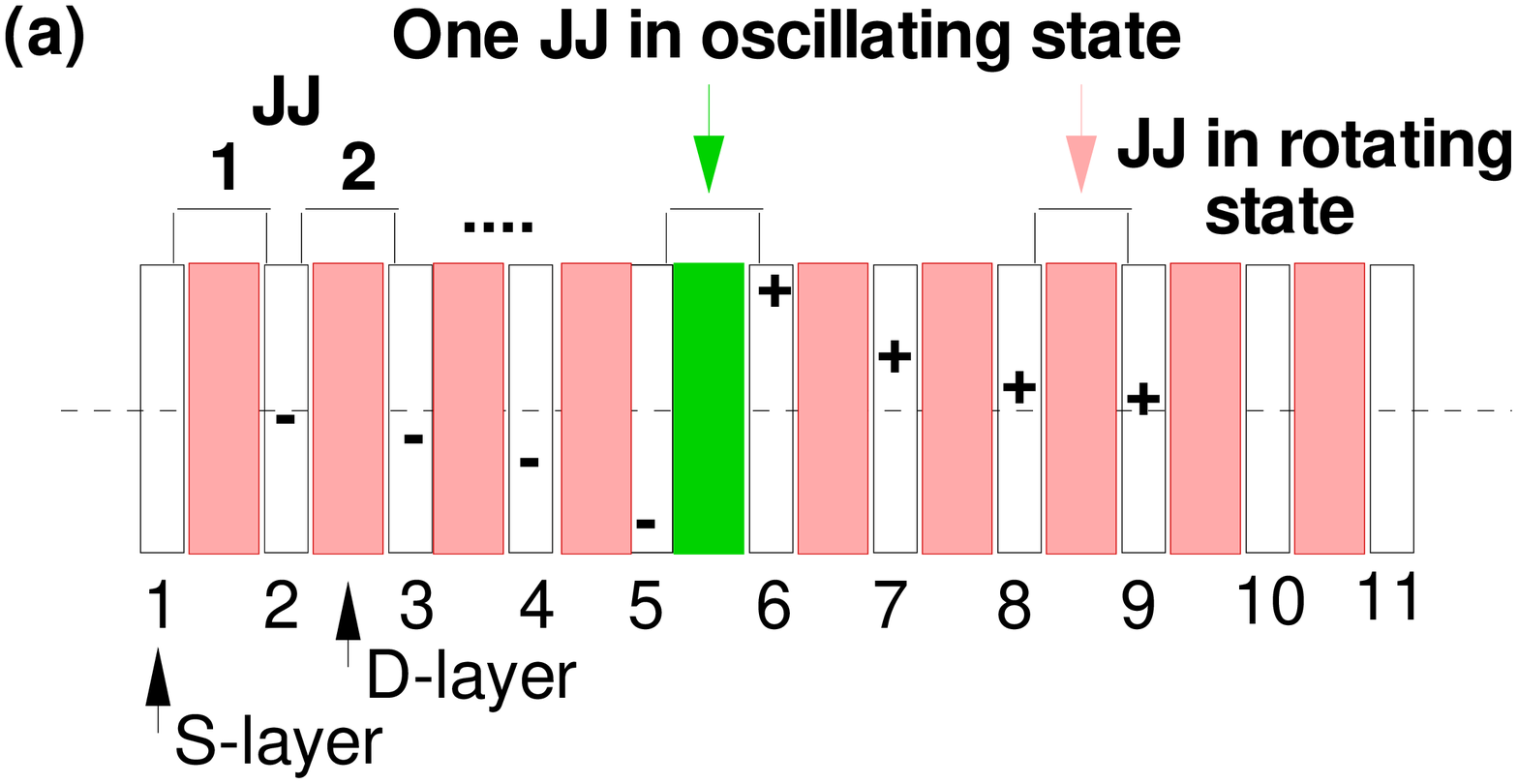}
\includegraphics[height=60mm]{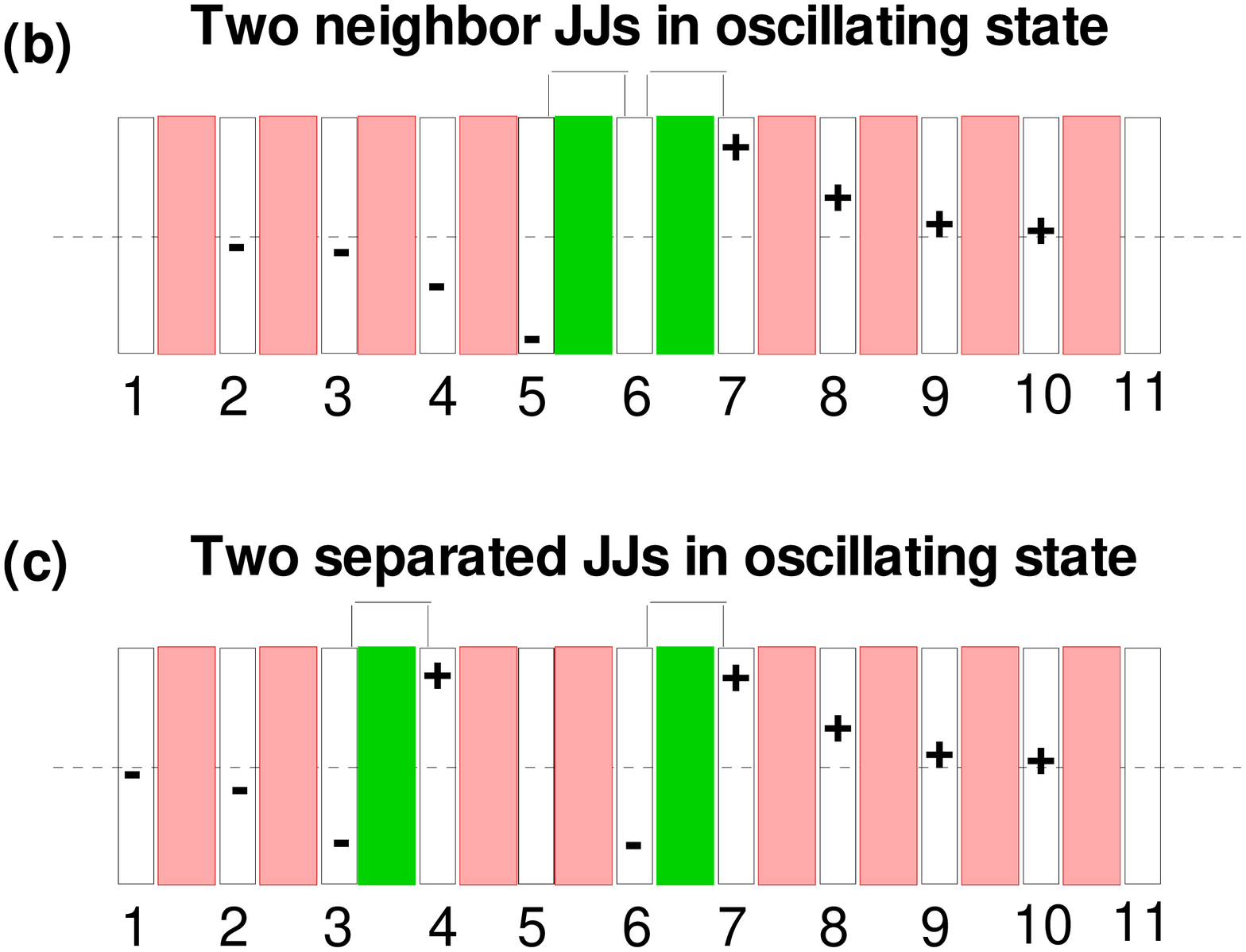}
\caption{(Color online) Modeling of charge distribution along the stack: (a) one oscillating JJ; (b) two neighbor oscillating junctions; (b) two separated  oscillating junctions.}
 \label{3}
\end{figure}

To explain the structure of realized CDW,  we use the idea that an inner branch corresponds to the state with JJ in $R-$ and $O-$states \cite{matsumoto99}. First we discuss what we would  obtain if one JJ of the stack is in the O-state.  Let it be JJ between S-layers 5 and 6, so voltage $V_5$ is close to zero $V_5=0$. We reflect the qualitative picture only. This situation is presented in  Fig.~\ref{3}(a). Because the charge in the 5-th S-layer is proportional to the voltage difference between neighboring JJ, $(V_5-V_4)$, its value is negative. By the same reasoning the charge in layer 6 is positive. As a result, we expect the CDW with the shape, shown in Fig.~\ref{3}(a) by pluses  and minuses. It counts the charge value relative to the dashed line.

\begin{figure}[h!t]
 \centering
\includegraphics[height=60mm]{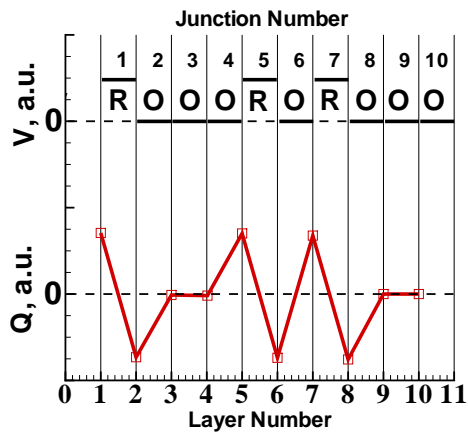}
\caption{(Color online) Schematic explanation of the CDW origin: (up) distribution of voltage among JJ; (down) distribution of charge among the S-layers.}
 \label{4}
\end{figure}

\begin{figure}
\includegraphics[height=85mm]{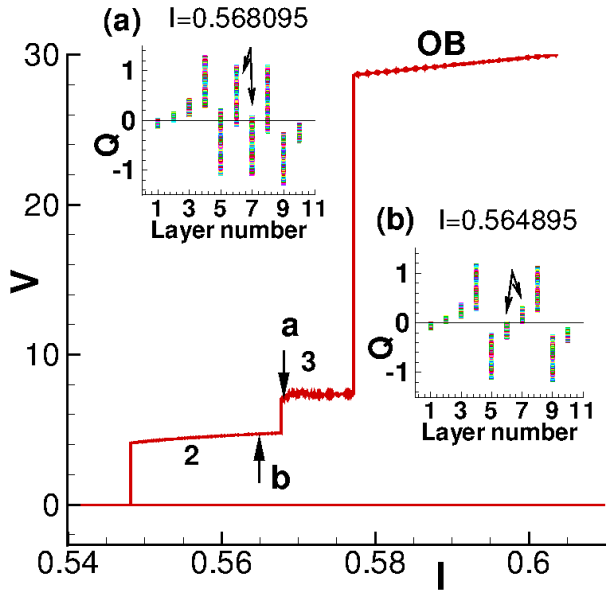}
\caption[bch!]{(Color online) IV-characteristics of the stack with 10 JJ with transition from branch number 3 to branch number 2. Inset (a) shows  CDW before transition at $I/I_c=0.568095$, while inset (b) shows CDW after transition to branch 2 at $I/I_c=0.564895$. Arrows stress the change of the charge variation  in S-layers 6 and 7.}
\label{5}
\end{figure}

\begin{figure}
\includegraphics[height=65mm]{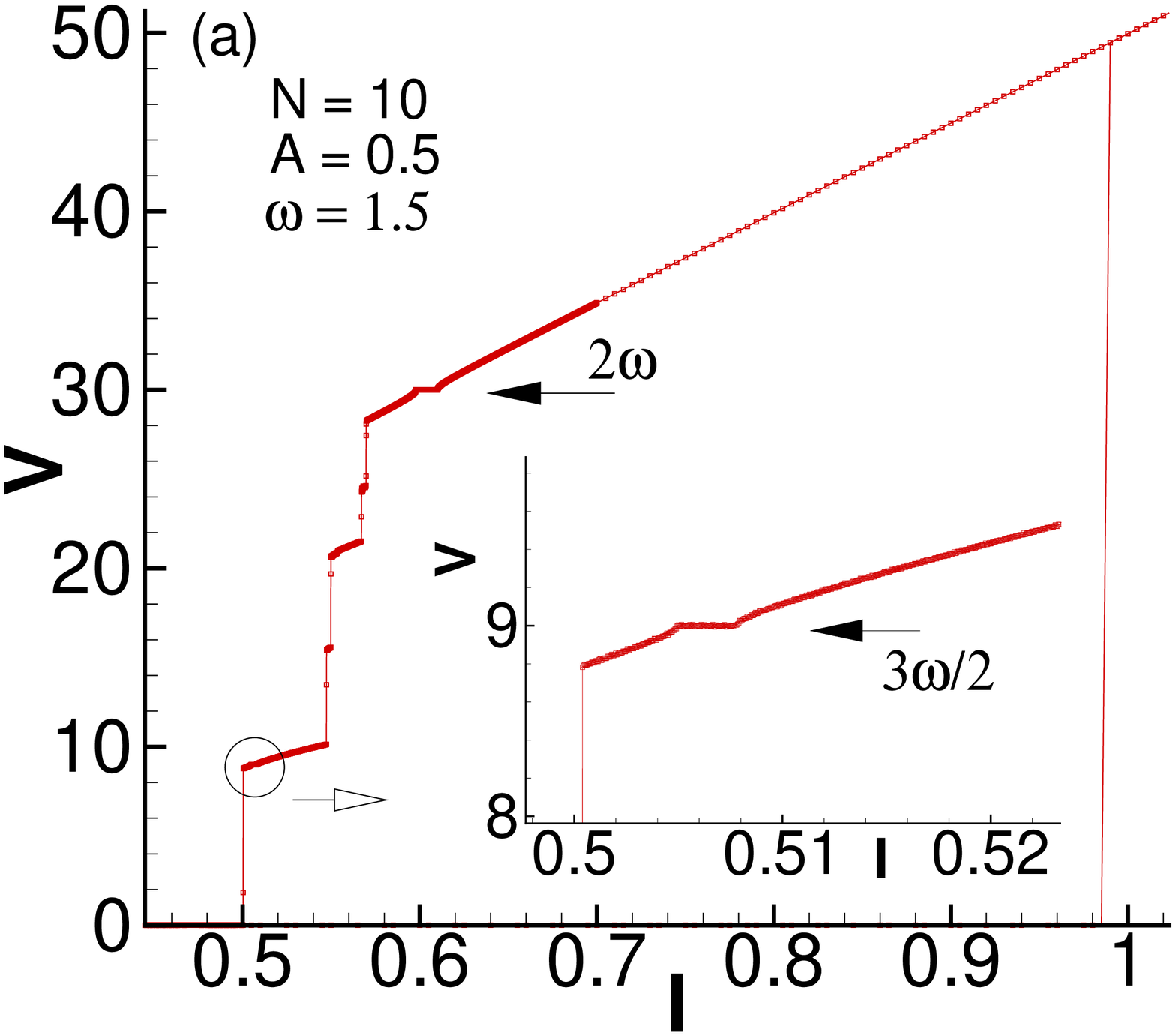}
\includegraphics[height=65mm]{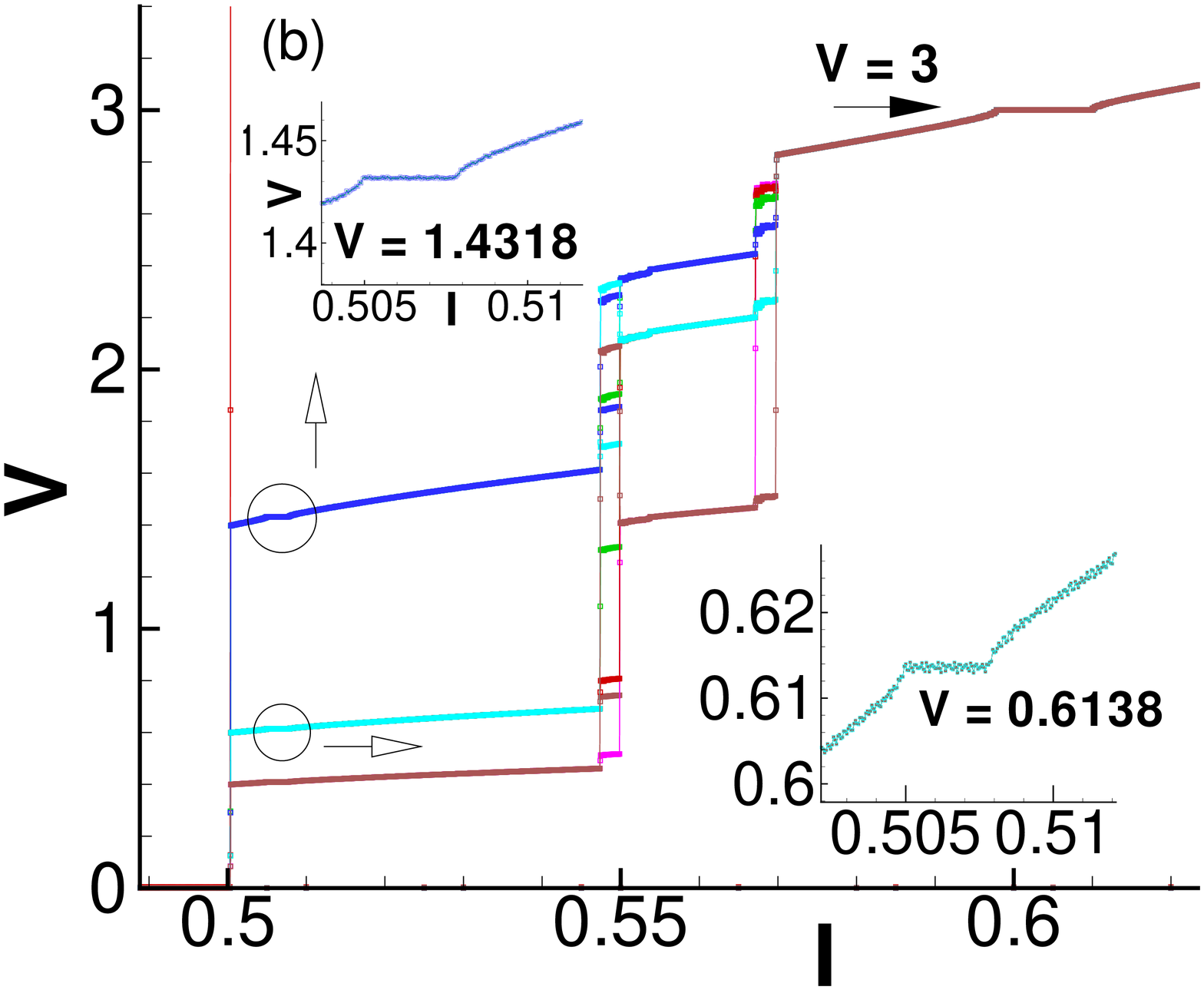}
\includegraphics[height=65mm]{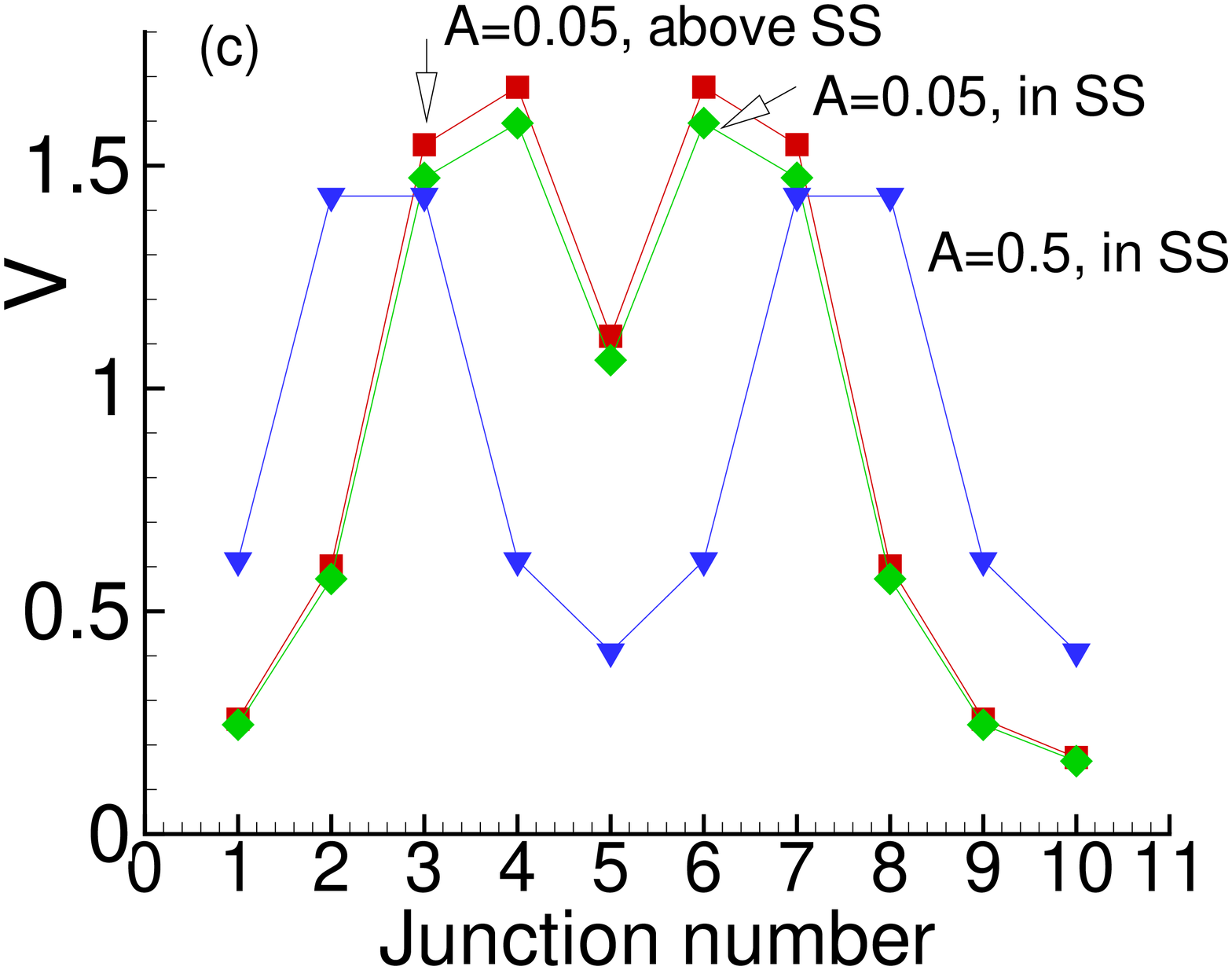}
\caption[]{(Color online) (a) IV-characteristics of the stack with 10 JJ under irradiation with frequency $\omega=1.5$ and amplitude $A=0.5$. Inset demonstrates the Shapiro step in the inner branch; (b) IV-characteristics of each JJ of the stack. Insets enlarges the parts with the Shapiro steps in two branches; (c) Distribution of  voltage along the stack at $A=0.5$ at the Shapiro step (gradient) and at $A=0.05$ above the Shapiro step (square) and at the Shapiro step  (diamond).}
\label{6}
\end{figure}

The two cases with two JJ in the O-state, neighbor and separated, with corresponding CDWs, are presented in Fig.~\ref{3}(b) and Fig.~\ref{3}(c), respectively.
Using this modeling, we can easily explain the results of Fig.~\ref{2}(d) and reproduce the created CDW. Because the charge's sign in  the layers 1 and 2, 5 and 6, and 7 and 8 is changing from plus to minus, the corresponding JJs (1,5, and 7) are in the rotating state as shown in the upper part of  Fig.~\ref{4}. Taking into account that charge is proportional to the difference in voltage between neighboring junctions, we find that the CDW has the structure shown in Fig.~\ref{4}, in agreement with the results in Fig.~\ref{2}(d).

Let us now consider the transformation from one CDW to another, which is  related to transitions between inner branches of the IV-characteristics. Fig.~\ref{5} shows the IV-characteristics for the following transitions: from outermost branch  $OB\rightarrow$ branch $B3\rightarrow$ branch $B2\rightarrow$ zero voltage state.  The corresponding CDWs are shown in the insets. Transition  $3\rightarrow 2$ corresponds to the transformation of one JJ from the $R$-state to the $O$-state. Comparing the CDWs shown in insets (a) and (b), we conclude that an additional transition has occurred between layers 6 and 7, as shown by the arrows in the insets.  We note that the state ($OOOORRROOO$), which  corresponds to the branch 3 in Fig.~\ref{5}, is  different from the state ($ROOOROROOO$) realized at the transition shown in Fig.~\ref{1}(c).

Let us now discuss the effect of the external radiation on  CDW corresponding to the inner branch. Figure ~\ref{6}(a) shows the IV-characteristics of the stack with 10 JJ under irradiation with frequency $\omega=1.5$ and amplitude $A=0.5$. We see the  Shapiro step in the outermost branch at $V=30$, which corresponds to the second harmonic $2\omega$. The IV-characteristic manifests four inner branches; their analysis shows that the lowest one corresponds to the state $ORROOORROO$  with four junctions in the rotating state.   The inset enlarges the circled part on this branch, where we see the step  at $V=9$ corresponding to the $3\omega/2$  Shapiro harmonic. Indeed, $V=N_R\frac{m}{n}\omega$ at  $N_R=4$, $m=3$ and $n=2$.

If all JJ in the stack were independent and reacted in the same way on the external radiation, we would observe the steps at $V=2.25$ in the IV-characteristics of all these four rotating JJ . However, as we see in the  insets to Fig.~\ref{6}(b), the Shapiro steps  appear at absolutely different values. We note that Fig.~\ref{6}(b) demonstrates the IV-characteristics for all JJ in the stack. We see that some parts of these characteristics for different junctions are coincide. The distribution of the Shapiro step voltage at this frequency $\omega=1.5$ and $A=0.5$ is shown in Fig.~\ref{6}(c). Four JJ show steps at $V=1.4318$, the other four at $V=0.6138$ and two JJ show steps at $V=0.4093$. These values are shown by gradients. The sum of these voltages in all JJ is equal to $V=9$. \emph{This fact indicates that the system of Josephson junctions behaves like a single whole system}.

To stress this phenomenon, we show in Fig.~\ref{6}(c)  additionally the distribution of junction voltage at the same frequency but with a smaller  amplitude $A=0.05$. In this case, the branch with four rotating junctions is  also realized in the total IV-characteristic of the stack but with another distribution $ORROOORROO$  of rotating and oscillation junctions. This branch  also demonstrates the Shapiro step at $V=9$. Squares and diamonds show the distribution of voltage along the stack:  above the Shapiro step and at the Shapiro step , respectively (IV-characteristic does not show here). Again, we observed the distribution of the Shapiro step voltage, but the sum of the voltages in all JJ is equal to $V=9$.

In the case when all JJ of the stack are in the rotating state (outermost branch), the Shapiro steps are realized at the same value of voltage in each IV-characteristic of the corresponding JJ, i.e., in a usual way.  We see in Fig.~\ref{6}(b) that all IV-curves of JJ in the stack demonstrate the Shapiro steps at the same value of voltage $V=3$. So we stress once again: when one or some JJs of the stack turn into the O-state, we observe the  system of the Shapiro steps at different voltages. The radiation frequency "manifests itself directly" in the IV-characteristics of the whole  stack only,  i.e., the system of coupled Josephson junctions behaves like an entire one.

Let us discuss shortly the possibility of the experimental testing of the obtained results. As we mentioned above, the charge $Q_l$  in the S-layer $l$ is proportional to the difference between the voltages $V_{l}$ and $V_{l+1}$ in the neighbor insulating layers $Q_l=Q_0 \alpha (V_{l+1}-V_{l})$,
where $Q_0 = \varepsilon \varepsilon _0 V_0/r_D^2$. For $r_D=3\times10^{-10}m$, $\varepsilon=25$, $\omega_p=10^{12}s^{-1}$ we get $V_0=3\times10^{-4} V$ and $Q_0=8\times10^5 C/m^3$. So at $Q=Q_0$ for a superconducting layer with the area $S=1 \mu m^2$ and thickness $d=3\times10^{-10} m$ the charge value is about $2.4\times10^{-16}C$. This value of charge is not high but it creates novel interesting physics and can be measured experimentally.

To summarize, we note that the use of intrinsic JJ as elements for superconductive electronics requires the knowledge of mechanisms of IV-characteristics branching and switching between states with different numbers of oscillating and rotating Josephson junctions. We demonstrated that in the system of coupled JJs within the hysteretic region, a transition  from the outermost to the inner branch is caused by the transformation of a LPW to a breathing CDW. At the transition from the outermost branch to the inner branch we also found a formation of an evanescent CDW or LPW from the states attributed to the outermost branch (with all junctions in the rotating state). The breathing CDW is a characteristic of the chosen inner branch and it is specified by the number of JJs  in the rotating and oscillating states, as well as  their positions in the stack.  Transitions between inner branches are caused by the transformation of one type of CDW to another one. The effect of external electromagnetic radiation on the system of coupled Josephson junctions in the CDW state is completely different from the case of single JJ. It causes the appearance of the set of the Shapiro steps in the IV-characteristics of JJ of the stack related to the voltage distribution among JJs. However, usual harmonics and subharmonics of radiation frequency are observed in the total IV-characteristics of the stack.

We thank I. R. Rahmonov, M. Gaafar, A. E. Botha, M. R. Kolahchi, P. Seidel for fruitful discussions. We acknowledge the support from JINR-EGYPT collaboration.

\bibliographystyle{apsrev}
\bibliography{CDW}

\end{document}